\documentclass[10pt,twocolumn,letterpaper]{article}

\usepackage[T1]{fontenc}
\usepackage{lmodern}           % scalable Latin Modern Type 1 (no Type 3 bitmap fallback)
\usepackage[utf8]{inputenc}
\usepackage{textcomp}
\usepackage[letterpaper,top=0.75in,bottom=0.9in,left=0.68in,right=0.68in,columnsep=0.25in]{geometry}
\usepackage{graphicx}
\usepackage{booktabs}
\usepackage{array}
\usepackage{amsmath}
\usepackage{caption}
\usepackage{titlesec}
\usepackage{xcolor}
\usepackage[hidelinks]{hyperref}

\graphicspath{{figures/}}
\renewcommand{\thesection}{\Roman{section}}
\renewcommand{\thesubsection}{\Alph{subsection}}
\titleformat{\section}{\normalfont\normalsize\scshape\centering}{\thesection.}{0.6em}{}
\titlespacing*{\section}{0pt}{1.1\baselineskip}{0.5\baselineskip}
\titleformat{\subsection}{\normalfont\normalsize\itshape}{\thesubsection.}{0.5em}{}
\titlespacing*{\subsection}{0pt}{0.6\baselineskip}{0.25\baselineskip}

\begin{document}

\twocolumn[{%
  \centering
  {\LARGE\bfseries Open, Reproducible Per-Bidding-Zone Carbon Intensity\\[2pt]
   for Nordic Power Systems\par}
  \vspace{0.9em}
  {\large Eirik Botten Nicolaysen\par}
  \vspace{0.2em}
  {\normalsize EcoDeco AS, Norway\par}
  {\normalsize ORCID:~0009-0001-9188-6788 \quad \href{mailto:eirik@ecodeco.no}{eirik@ecodeco.no}\par}
  \vspace{1.4em}
}]

\noindent\textbf{\textit{Abstract}---}%
\textbf{Carbon-aware scheduling of electrical loads depends on accurate grid carbon-intensity (CI) signals, yet existing tools commonly apply a uniform placeholder emission factor that misrepresents regional variation. This work presents an open, reproducible method for computing per-bidding-zone CI across the Nordic power system. Generation-by-type data from ENTSO-E is combined with IPCC AR5 life-cycle emission factors, duration-weighted per bidding zone, with an explicit materiality threshold for missing data. Applied to nine bidding zones (NO1--NO5, SE1--SE4) for 2025, the method yields CI values ranging from 14.5 to 39.6~gCO\textsubscript{2}eq/kWh. The results reveal that the uniform 18.0~gCO\textsubscript{2}eq/kWh placeholder is wrong in both directions: it understates CI for most zones, by a factor of approximately 2.2 for NO4, yet overstates CI for the nuclear-dominated zone SE3, whose production-based intensity is the lowest of the nine. Drift is characterised against a pre-registered threshold of 15\% year-on-year change, including a borderline case in SE4 and event-driven variation in NO4. Forecasting experiments do not reject the null hypothesis that simple baselines suffice, in either the Norwegian or Swedish zones. The full method, data, and results are released under open licences and archived at DOI 10.5281/zenodo.21042581, with the per-zone values contributed upstream to the codecarbon library. Per-bidding-zone resolution is necessary for accurate carbon accounting in Nordic grids, where country-level averages obscure substantial intra-country variation.}

\vspace{0.6em}
\vspace{0.8em}

\section{Introduction}

Carbon-aware computing---shifting flexible computational load in time or space toward intervals when the electricity supply is cleaner---has matured from a proposal into an established practice, with simulation frameworks, scheduling policies, and a substantial ``wait-awhile'' literature now in regular use [1], [2]. Every such technique rests on the same input: a time-resolved estimate of the carbon intensity (CI) of the electricity a workload would consume, expressed in gCO\textsubscript{2}eq/kWh. The accuracy and spatial resolution of that grid signal therefore bound the accuracy of any scheduling decision built on top of it. From a power-systems standpoint this is a load-management problem---aligning demand with the instantaneous emissions profile of generation---and its effectiveness depends directly on how faithfully the CI signal reflects the actual zonal generation mix that serves the load.

For the Nordic electricity system this signal is poorly served, despite the region's importance as a destination for energy-intensive computation. Existing openly available CI sources for Norway and Sweden fall into four categories, none of which provides a versioned, citable per--bidding-zone dataset. The first is uniform static factors: the widely used emissions-accounting tool codecarbon [3] assigns a single constant of 18.0~gCO\textsubscript{2}eq/kWh to every Norwegian and Swedish bidding zone alike (verified against the installed distribution). The second is country-level aggregation: the established CI-forecasting systems CarbonCast [4] and EnsembleCI [5] cover Sweden only as a single national series, with no per-zone decomposition, and Norway is absent from them altogether. The third is the peer-reviewed literature: studies that do report per-zone Nordic values, discussed below, but do not deposit the underlying series under a persistent identifier, so a specific published value cannot be cited and recomputed against a fixed version. The distinction is deposition rather than licensing: the factor bases these studies rest on are in part openly available. The fourth is operational platform data. Electricity Maps publishes hourly carbon intensity for all nine Nordic bidding zones, computed by a flow-tracing method published in the peer-reviewed literature [7] and IPCC 2014 lifecycle factors, with open-source parsers; it is therefore auditable and citable as a service. What it is not is a versioned artifact archived under a persistent identifier, recomputable end to end from a published pipeline---the historical series sits behind a paid API, to be re-queried rather than reproduced. WattTime is likewise a commercial, access-gated service.

Open, reproducible carbon-intensity data for European power systems does exist---but at national resolution. On the production side, Unnewehr et al. [6] derive per-country, per-technology emission factors entirely from open ENTSO-E and emissions-trading data, demonstrating that a production-based CI signal is attainable at the country level from open sources alone. On the consumption side, Tranberg et al. [7] introduce the flow-tracing method underlying the Electricity Maps service, attributing emissions from producer to consumer across the coupled European market, and Zhang et al. [8] release ECON-PowerCI, a near-real-time consumption-based daily dataset covering 35 European countries. All three stop at the national aggregate; none resolves the electricity bidding zone. Yet the bidding zone is where the Nordic system is organized: Norway and Sweden are each split into several zones (NO1--NO5, SE1--SE4) whose generation mixes---and therefore carbon intensities---differ materially within a single country. Per--bidding-zone resolution is in any case already established in this market: Phan et al. [11] forecast electricity prices across Norway's five zones individually. The zone, not the country, is the unit the Nordic system is settled and operated in. Per--bidding-zone carbon intensity for the Nordic zones is not new. Clau{\ss} et al. [12] report hourly average intensities for all five Norwegian zones, import-adjusted and production-side, for 2015; Engstam et al. [13] and Papageorgiou et al. [14] report per-zone Swedish factors for 2018--2021 and 2018 respectively. Our contribution is therefore not one of priority. It is that the series is archived under a persistent identifier, versioned, and recomputable end to end from a published pipeline on a factor base that is free to obtain---IPCC AR5 Annex III, as against Ecoinvent in Clau{\ss} et al. and Papageorgiou et al., elmada in Engstam et al., and IPCC 2014 in Electricity Maps---and that it is current, covering 2021--2025 rather than a single earlier year.

The cost of the gap is not merely coarse resolution but quantifiable error. The uniform 18.0 placeholder is wrong for every Nordic zone, and wrong in both directions at once: it underestimates the gas-influenced zone NO4 by roughly a factor of 2.2 and overestimates the nuclear-dominated zone SE3, whose production-based intensity lies well below the constant. A single scalar cannot repair this, because the bias has opposite sign depending on the zonal generation mix---an error structure that only per-zone values can resolve. We base this on our own measurements against the installed placeholder, reported in Section~III; we are aware of no external study reporting a single headline error figure for these factors, and we therefore make no such external attribution here.

We present Khepri, an open and reproducible per--bidding-zone CI dataset and method for the nine Nordic bidding zones---NO1--NO5 and SE1--SE4. Openness and reproducibility at the national level are already established by the work above, and per-zone Nordic intensities are already published; Khepri's distinct contribution is to carry openness, versioning and independent recomputability to that resolution. Our contribution is fourfold. First, we derive production-based CI for each zone deterministically from ENTSO-E generation-by-type data [9] and published IPCC AR5 lifecycle emission factors [10], with every methodological degree of freedom fixed in a public chain of architecture decision records committed before computation. Second, we characterize the signal's behavior over time with pre-registered drift and forecast analyses---reporting, among other findings, the two-way placeholder error, a per-zone drift characterization that distinguishes genuine trends from one-off industrial events, and an honestly negative forecast result in which heavy machine learning does not beat a simple seasonal baseline. Third, we demonstrate adoption rather than leaving the dataset on the shelf, contributing the per-zone values upstream into codecarbon. Fourth, we extend the same method from Norway to the Swedish zones; we are explicit that the Swedish contribution is one of \textit{granularity}---an archived, recomputable per-zone series rather than a first---and not a claim of a stronger or more distinct signal than the Norwegian track, whose cross-zone variation is in fact structurally wider. We bound the contribution honestly: the method is production-based and does not compute consumption-based, flow-traced, or marginal emissions, which we treat as a separately scoped extension rather than an implied capability.

The remainder of this paper is organized as follows. Section~II details the data source, emission factors, CI computation, and the pre-registered drift and forecast methods; Section~III reports the per-zone intensities, the placeholder error, the drift characterization, and the forecast outcome; Section~IV discusses the adoption pathway and limitations; and Section~V concludes.

\section{Methodology}

\subsection{Data Source}

We derive per--bidding-zone carbon intensity (CI, gCO\textsubscript{2}eq/kWh) for the nine Nordic electricity bidding zones NO1--NO5 (Norway) and SE1--SE4 (Sweden) from the ENTSO-E Transparency Platform's \textit{Actual Generation per Production Type} dataset (document type A75) [9], queried per zone at native temporal resolution. Each zone is addressed by its individual EIC code rather than the country aggregate; the Swedish per-zone codes (SE1 \texttt{10Y1001A1001A44P}, SE2 \texttt{\ldots A45N}, SE3 \texttt{\ldots A46L}, SE4 \texttt{\ldots A47J}) were verified against the \texttt{entsoe-py} area mapping, and the country aggregate code \texttt{10YSE-1-\/-\/-\/-\/-\/-\/-\/-K} was explicitly identified and rejected, since using it would collapse the per-zone distinction this work is built on. The headline CI figures are computed for calendar year 2025. The same pipeline is then applied across 2021--2025 for Norway and 2022--2025 for Sweden to characterize drift; the Swedish 2021 year is excluded because its coverage is only 4.7\% (approximately two weeks of reported data), an exclusion fixed before analysis rather than after inspecting results.

We emphasize at the outset that the signal is \textit{production-based}: CI reflects the generation mix physically produced within a zone, weighted by lifecycle emission factors. Import and export flows are deliberately excluded. Consumption-based or flow-traced CI, and marginal-emissions accounting, are distinct quantities that this method does not compute; we treat them as a separately scoped extension (Section~IV-C) rather than claiming them here.

\subsection{Emission Factors}

Generation is converted to CI using lifecycle emission factors from IPCC WG III AR5, Annex III (lifecycle medians) [10]. These are \textit{lifecycle} values---encompassing construction, fuel cycle, operation, and decommissioning---not direct combustion-only factors, which makes the resulting intensities comparable with the lifecycle convention used in the carbon-aware computing literature rather than with plant-stack emissions alone. Because the IPCC primary PDF was not directly reachable from the build environment, the values were taken from a checkable secondary source that names the table explicitly, and the two highest-leverage factors (coal = 820, gas = 490) were independently confirmed against the IPCC document's own PDF. Each factor is annotated in the source code with its mapping and provenance.

The factors applied to the production types that occur materially in the Nordic data are: Fossil Gas = 490; all hydro categories (Water Reservoir, Run-of-river and poundage, and Pumped Storage) = 24; Wind Onshore = 11; Wind Offshore = 12; Solar = 48; and Nuclear = 12. Nuclear is the one production type present in Sweden but absent from Norway, and within Sweden it is observed empirically only in SE3; its factor (12, the AR5 lifecycle median) was cross-checked against the same secondary source. One mapping is flagged honestly rather than presented as exact: Hydro Pumped Storage is assigned the hydro factor (24) as a \textit{proxy}, since its true footprint depends on the charging source. Coal (820), geothermal (38), and biomass (230) are likewise retained for robustness but do not occur materially in these zones; biomass appears only as an all-zero column (NO3) or at a negligible annual mean of 0.012~MW (NO1)---both well below the materiality threshold---and is absent altogether elsewhere. Fossil Oil, which does not appear in the Nordic data, carries a flagged approximation (650) used only for robustness should it appear.

\subsection{Carbon-Intensity Computation}

For each reporting interval, the instantaneous CI is the generation-weighted mean of the per-type factors,
\begin{equation}
\mathrm{CI} = \frac{\sum_t \mathrm{MW}_t \cdot f_t}{\sum_t \mathrm{MW}_t},
\end{equation}
summed over production types $t$. Because the 2025 data arrives at mixed resolution (both 15-minute and 60-minute periods, verified on disk), we aggregate to a zone-year figure by weighting each interval by its \textit{duration} in hours rather than by interval count:
\begin{equation}
\overline{\mathrm{CI}} = \frac{\sum_t \mathrm{MW}_t \cdot f_t \cdot \Delta h}{\sum_t \mathrm{MW}_t \cdot \Delta h}.
\end{equation}
This is equivalent to total emissions divided by total energy over the clean intervals, and it avoids the distortion that count-weighting would introduce when 15- and 60-minute periods are mixed. The computation is deterministic and fully reproducible from the raw ENTSO-E extract and the published method.

\subsection{Missing-Data Handling and Materiality Threshold}

ENTSO-E series contain NaN entries where a production type is unreported for an interval. We treat NaN as genuinely missing, not as zero, since silently zero-filling would bias the average downward whenever real generation is unreported. An interval is therefore excluded from the average if a \textit{material} production type carries NaN. To prevent this rule from discarding otherwise valid data on account of trivially small types, we pre-register a materiality threshold: a type is \textit{negligible} in a zone if its annual-mean contribution is below 0.5\% of the zone's mix \textbf{or} below 5~MW in absolute terms; NaN in a negligible type is treated as 0, while an interval is excluded only when NaN occurs in a material type. This threshold was set on principled grounds---a type below it cannot move the energy-weighted average by more than a fraction of a gCO\textsubscript{2}eq/kWh regardless of its factor---and explicitly \textit{not} tuned against the resulting coverage. It was introduced after an initial run revealed a coverage artefact (NO2 and NO3 coverage falling to roughly 31\% and 27\%, driven entirely by minor unreported types such as offshore wind at an annual mean of 2.5~MW); fixing the artefact required a principled threshold rather than a discretionary one, and we report per-zone coverage and the negligible-type classification as provenance. Production types with no verified factor in the chosen source (Waste, Other, Other renewable) are excluded from the primary CI and reported only as a sensitivity, never folded into the headline number.

\subsection{Drift Characterization}

A CI figure is only useful to a downstream tool if its shelf life is known. We therefore characterize year-over-year drift in the production-based signal using a threshold \textit{pre-registered before any drift result was observed}: a zone exhibits material drift if its year-over-year CI change exceeds 15\% \textbf{or} if the mix share of a material type shifts by more than 5 percentage points. The pre-registration is itself a methodological commitment, not a presentational one: the threshold's provenance is the Norwegian analysis (ADR-0003), locked before any Swedish data existed, and it is applied to the Swedish zones without adjustment---borderline cases just above the threshold are reported as material drift rather than smoothed away because they are close. The associated null hypothesis is stated explicitly and symmetrically: H\textsubscript{0} holds that hydro-dominated per-zone CI is stable year over year and that a prior-year figure is a good proxy. We do not assume an outcome; a stable signal and a drifting signal are both genuine findings with different consequences for how often a downstream integration must refresh its values. The regime comparison between the 2021--2022 energy-crisis years and the later years is reported as an effect size against the threshold, not as a \textit{p}-value, because the data constitute the full population of intervals rather than a sample.

\subsection{Forecasting Setup}

To assess whether the per-zone signal supports multi-day carbon-aware scheduling, we produce 96-hour, day-wise forecasts (day 1 = 0--24~h through day 4 = 72--96~h) at hourly resolution, matching the evaluation convention of the carbon-aware forecasting literature for comparability. Model complexity is escalated in a locked low-to-high order. Persistence (flat and diurnal) serves as a floor; a Seasonal-ARIMA (SARIMA) model is the primary baseline, as it is the field's standard benchmark; and a gradient-boosting machine (GBM) is introduced only if the simpler baseline is documented as insufficient and the improvement is material. The forecast accuracy metric is MAPE, supplemented---by pre-registration, not post hoc---with MAE and RMSE (because the low Nordic CI regime makes MAPE unstable) and with a concordance index, which captures whether the forecast ranks clean and dirty hours correctly, the property that actually matters for scheduling. The pre-registered H\textsubscript{0} states that a heavy ML model does not meaningfully beat the SARIMA/persistence baseline; a negative result---simple models suffice---is treated as an equally publishable finding with a direct adoption consequence. For Norway we use a double split: a drift-aware primary split (train 2021--2023, validate 2024, test 2025) reported separately for stable versus drifting zones, and a field-exact secondary split (train 2019--H1 2021, test H2 2021) that yields one directly comparable figure against the field's published low-CI hydropower results. For Sweden we use only the drift-aware split (train 2022--2023, test 2025); we state plainly that the field-exact secondary split is \textit{not} reproduced for Sweden, because there is no pre-2022 Swedish data with adequate coverage and no per-zone external reference to lock against (existing field coverage of Sweden is country-aggregate only), so a secondary split would add a period without adding comparability.

Throughout, the Swedish results are positioned as a \textit{granularity} contribution---an archived, recomputable per-zone series---rather than as a stronger or richer signal than the Norwegian track. The cross-zone variation in Sweden is structurally narrower than in Norway, and we make no magnitude claim for either track over the other; the contribution is the archived, versioned series on a freely available factor base, not the existence of per-zone Swedish factors, which Engstam et al. [13] and Papageorgiou et al. [14] already report.

\subsection{Use of Large Language Models}

Large language model--based tools were used to assist with code development, data-pipeline verification, and manuscript preparation; all method choices, numerical results, and conclusions were authored, verified against primary sources, and are the responsibility of the author.

\subsection{Reproducibility}

The method is reproducible end to end. Every degree of freedom is fixed in a public chain of Architecture Decision Records (ADR-0001 through ADR-0009) committed \textit{before} the corresponding computation, so the figures are verifiable rather than post-rationalized. The code is released under Apache-2.0 and the data and documentation under CC-BY-4.0; raw ENTSO-E extracts are not redistributed but are fetched reproducibly through the documented API queries. The artifact is archived with a persistent identifier, DOI 10.5281/zenodo.21042581.

\section{Results}

We report four results, in order: the per--bidding-zone carbon-intensity figures themselves (Table~\ref{tab:ci}); the magnitude and direction of the error in the uniform placeholder these figures replace (Table~\ref{tab:err}); the year-over-year drift of the signal against a pre-registered threshold (Table~\ref{tab:drift}); and the forecast outcome, which is an honest negative result for heavy machine learning. Throughout, the Swedish track is a granularity contribution---an archived, recomputable per-zone series---and not a claim of a stronger or richer signal than the Norwegian track.

\subsection{Per--Bidding-Zone Carbon Intensity}

Table~\ref{tab:ci} gives the production-based CI for all nine zones in calendar year 2025. The figures are low in absolute terms, as expected for hydro-dominated systems, but they are not uniform: they span a factor of 2.7, from 14.53~gCO\textsubscript{2}eq/kWh in SE3 to 39.65 in NO4. The two extremes are mechanistically distinct. SE3 is the lowest of the nine because it is nuclear-dominated (nuclear at the AR5 lifecycle factor of 12, observed only in SE3 among the Swedish zones), while NO4 is the highest because it carries fossil gas year-round from the Hammerfest LNG (Melk\o ya) plant. The four pure-hydro Norwegian zones (NO1, NO2, NO3, NO5) cluster tightly between 21.46 and 24.46, confirming that, absent gas or nuclear, the production-based signal is genuinely flat across hydro zones---a point we return to in the forecast results.

\begin{table}[t]
\centering
\caption{Per--bidding-zone production-based carbon intensity, 2025 (gCO\textsubscript{2}eq/kWh).}
\label{tab:ci}
\small
\begin{tabular}{@{}l r >{\raggedright\arraybackslash}p{3.6cm}@{}}
\toprule
Zone & CI 2025 & Dominant driver \\
\midrule
NO1 & 23.31 & hydro \\
NO2 & 23.85 & hydro \\
NO3 & 21.46 & hydro \\
\textbf{NO4} & \textbf{39.65} & fossil gas (Hammerfest LNG) --- \textbf{highest} \\
NO5 & 24.46 & hydro \\
SE1 & 20.63 & hydro / wind \\
SE2 & 20.11 & hydro / wind \\
\textbf{SE3} & \textbf{14.53} & nuclear-dominant --- \textbf{lowest} \\
SE4 & 17.42 & wind / solar \\
\bottomrule
\end{tabular}
\end{table}

\subsection{The Two-Way Placeholder Error}

These nine distinct values replace a single uniform constant. The carbon-accounting tool codecarbon assigns 18.0~gCO\textsubscript{2}eq/kWh to every Norwegian and Swedish bidding zone alike. Table~\ref{tab:err} shows that this constant is wrong for every zone, and---importantly---wrong in \textit{both} directions. For seven of the nine zones it underestimates the production-based intensity, most severely for NO4, where the production-based 2025 value is roughly 2.2$\times$ the placeholder. For the two lowest-intensity zones it instead \textit{over}estimates: most materially for the nuclear-dominated SE3, whose production-based value (14.53) lies well below 18.0, and marginally for SE4 (17.42). A uniform constant therefore cannot be corrected by a single scalar adjustment: it is biased low where gas is present and high where nuclear dominates, and only per-zone values resolve this. We note that the per-zone deviation percentages in Table~\ref{tab:err} are computed from the verified 2025 values of Table~\ref{tab:ci}; they are not pre-computed on disk.

\begin{table*}[t]
\centering
\caption{Error in the uniform 18.0~gCO\textsubscript{2}eq/kWh placeholder, by zone.}
\label{tab:err}
\small
\begin{tabular}{@{}l r r l r@{}}
\toprule
Zone & Production-based CI 2025 & Placeholder & Direction & Deviation\textsuperscript{\dag} \\
\midrule
NO1 & 23.31 & 18.0 & underestimate & $-22.8\%$ \\
NO2 & 23.85 & 18.0 & underestimate & $-24.5\%$ \\
NO3 & 21.46 & 18.0 & underestimate & $-16.1\%$ \\
NO4 & 39.65 & 18.0 & underestimate & $-54.6\%$ ($\approx$2.2$\times$ too low) \\
NO5 & 24.46 & 18.0 & underestimate & $-26.4\%$ \\
SE1 & 20.63 & 18.0 & underestimate & $-12.8\%$ \\
SE2 & 20.11 & 18.0 & underestimate & $-10.5\%$ \\
SE3 & 14.53 & 18.0 & \textbf{overestimate} & $\mathbf{+23.9\%}$ \\
SE4 & 17.42 & 18.0 & overestimate & $+3.3\%$ \\
\bottomrule
\end{tabular}

\vspace{2pt}
{\footnotesize \textsuperscript{\dag}\,Deviation $=$ (placeholder $-$ production-based)$/$production-based, computed from the 2025 values in Table~\ref{tab:ci}; negative $=$ placeholder too low. The placeholder value 18.0 (year 2024) is verified in the installed codecarbon distribution.\par}
\end{table*}

\subsection{Drift Against a Pre-Registered Threshold}

Whether a one-year CI figure remains valid into the next year determines how often a downstream tool must refresh it. We test this against a threshold fixed before any drift result was observed: a zone shows material drift if its year-over-year CI change exceeds 15\% \textbf{or} if a material production type's mix share shifts by more than 5 percentage points. Table~\ref{tab:drift} reports the outcome over 2021--2025 for Norway and 2022--2025 for Sweden.

For Norway, the three pure-hydro zones (NO1, NO2, NO3) hold the stability null hypothesis comfortably---all year-over-year transitions stay below 5\%---so a prior-year figure is a sound annual proxy. NO4 and NO5 reject it, but for different reasons that we are careful to separate. NO5 is a genuine monotone trend: a real gas phase-out drives CI down from 34.90 (2021) toward the pure-hydro floor (24.46 in 2025). NO4 is \textit{not} a trend at all but an industrial event: the Hammerfest LNG plant was offline after a September 2020 fire and restarted on 2 June 2022, so NO4's large CI movements track an outage-and-recovery, not a smooth drift. We report NO4 as material drift against the threshold while stating plainly that its mechanism is an unpredictable event---a distinction that matters directly for the forecast in Section~III-D.

For Sweden, only SE2 clears both arms of the threshold and is stable on each. SE1 and SE3 hold their CI stable (changes of $-3.83\%$ and $+7.82\%$ over the window, both well under 15\%) yet register material \textit{mix-share} drift---SE1 through a hydro-to-wind shift, SE3 through a falling nuclear share---whose CI effect is offset by the relative factor values. SE4 is the one zone to cross the CI arm directly: its 2022$\to$2025 change is $+15.2\%$, just above the 15\% threshold. We report SE4 as material drift rather than rounding the 0.2 percentage-point margin away, in keeping with the pre-registration; the rise is monotone and solar-driven, with no V-shaped event signature. We flag a framing nuance for honesty: against the literal OR-threshold, SE1, SE3, and SE4 all register material drift (the first two on the mix arm with stable CI), whereas a CI-only reading would call SE1/SE2/SE3 stable. Both statements are true on different arms of the same pre-registered rule; we report the full picture rather than the simpler half.

\begin{table*}[t]
\centering
\caption{Drift against the pre-registered threshold ($>$15\% YoY CI change OR $>$5~pp material mix-share shift). Window: NO 2021--2025, SE 2022--2025.}
\label{tab:drift}
\footnotesize
\begin{tabular}{@{}l r >{\centering\arraybackslash}p{1.1cm} >{\centering\arraybackslash}p{2.3cm} >{\centering\arraybackslash}p{1.9cm} >{\raggedright\arraybackslash}p{4.4cm}@{}}
\toprule
Zone & CI change over window & CI $>$15\%? & Material mix drift ($>$5~pp)? & H\textsubscript{0} (stable) & Mechanism \\
\midrule
NO1 & $-1.4\%$ & no & no & \textbf{holds} & --- \\
NO2 & $+3.2\%$ & no & no & \textbf{holds} & --- \\
NO3 & $-0.9\%$ & no & no & \textbf{holds} & --- \\
NO4 & event\textsuperscript{\ddag} & yes & yes (gas) & \textbf{rejected} & industrial event (outage/recovery) \\
NO5 & $-29.9\%$ & yes & yes (gas) & \textbf{rejected} & genuine gas phase-out (trend) \\
SE1 & $-3.83\%$ & no & yes (hydro$\downarrow$/wind$\uparrow$) & rejected (mix arm) & mix drift, CI stable \\
SE2 & $-2.36\%$ & no & no & \textbf{holds} & stable on both arms \\
SE3 & $+7.82\%$ & no & yes (nuclear$\downarrow$) & rejected (mix arm) & mix drift, CI stable \\
SE4 & $\mathbf{+15.2\%}$ & \textbf{yes} & yes (solar$\uparrow$) & \textbf{rejected} & monotone, solar-driven \\
\bottomrule
\end{tabular}

\vspace{2pt}
{\footnotesize \textsuperscript{\ddag}\,NO4's window-total is dominated by the 2021 outage low and is not a meaningful single figure; its drift is event-driven (year-over-year swings up to $+61.4\%$, with the plant offline in 2021 and restarted June 2022), reported as material drift but explicitly not a trend.\par}
\end{table*}

\subsection{Forecast: A Negative Result for Heavy Machine Learning}

The forecast asks whether the per-zone signal supports multi-day carbon-aware scheduling, and specifically whether a heavy model earns its complexity. The pre-registered null hypothesis---that a gradient-boosting machine (GBM) does not meaningfully beat a SARIMA/persistence baseline---is \textbf{not rejected in either track.} For Norway, simple models are hard to beat: the improvement of SARIMA or GBM over flat persistence is small for the stable hydro zones, and for NO3 flat persistence is itself the best model (MAPE 6.84). The heavy model is, moreover, not a safe default---it collapses on the volatile zone, with GBM reaching a MAPE of 35.65 on NO4 in 2025 against SARIMA's 7.24. NO4 is the hardest zone precisely because its defining movement is the Hammerfest event, which is not present in the CI history and cannot be learned from it; this is the forecast-side confirmation of the drift mechanism in Section~III-C, not a model deficiency.

For Sweden the outcome is the same and is stated explicitly on disk as ``H0 not rejected'': SARIMA beats GBM in all four zones, with GBM degrading the day-1 MAPE by $+1.0$, $+1.5$, $+4.3$, and $+3.6$ percentage points for SE1 through SE4 respectively. Escalation to a heavy model is therefore unwarranted. We do not dress this as a partial victory and we do not reframe it as a weakness: that a simple baseline suffices is itself a publishable finding with a concrete adoption consequence---a downstream integration can refresh per-zone CI with a lightweight forecaster (or, for the stable zones, an annual update) rather than maintaining a heavy retraining pipeline. The per-zone day-1 MAPE values are usable (SE3 3.38, SE2 6.75, SE1 7.85; SE4 higher at 16.06, reflecting its documented drift), and lie in the same order of magnitude as the only verified field reference: the SE \textit{country-aggregate} day-1 MAPE of 8.87\% that EnsembleCI [5] reports for CarbonCast in its Table~2---a comparison we always state with the per-zone-versus-aggregate caveat, never as ``beats CarbonCast.'' We also report a limitation surfaced by the concordance index rather than hidden by MAPE: in SE1 and SE2 concordance is only 0.581 and 0.596, barely above the 0.5 coin-flip line, meaning the forecast tracks the CI \textit{level} acceptably but ranks cleaner-versus-dirtier hours poorly---a real limit for scheduling in those two zones, in contrast to SE3 (0.842) and SE4 (0.822) where direction tracking is strong.

Finally, the cross-zone structure underscores why the Swedish contribution is granularity rather than magnitude. The annual cross-zone spread---the gap between the highest- and lowest-intensity zone within a given year---ranges 16.7--30.7~gCO\textsubscript{2}eq/kWh for Norway across the comparison window but only 5.6--8.0 for Sweden. (This spread is a per-year between-zone quantity and should not be confused with the absolute CI levels of Table~\ref{tab:ci}.) Sweden's narrower spread means its per-zone signal is structurally less distinct than Norway's; the contribution is an archived, recomputable per-zone SE1--SE4 series on a freely available factor base, not the existence of per-zone Swedish factors, which Engstam et al. [13] and Papageorgiou et al. [14] already report. Neither track is claimed to be the stronger of the two.

\begin{figure}[t]
\centering
\includegraphics[width=\columnwidth]{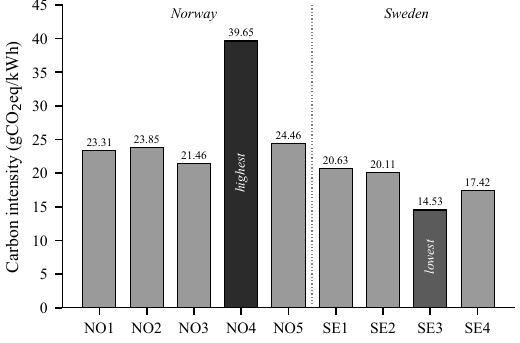}
\caption{Per-bidding-zone production-based carbon intensity, 2025.}
\label{fig:ci}
\end{figure}

\begin{figure*}[t]
\centering
\includegraphics[width=0.86\textwidth]{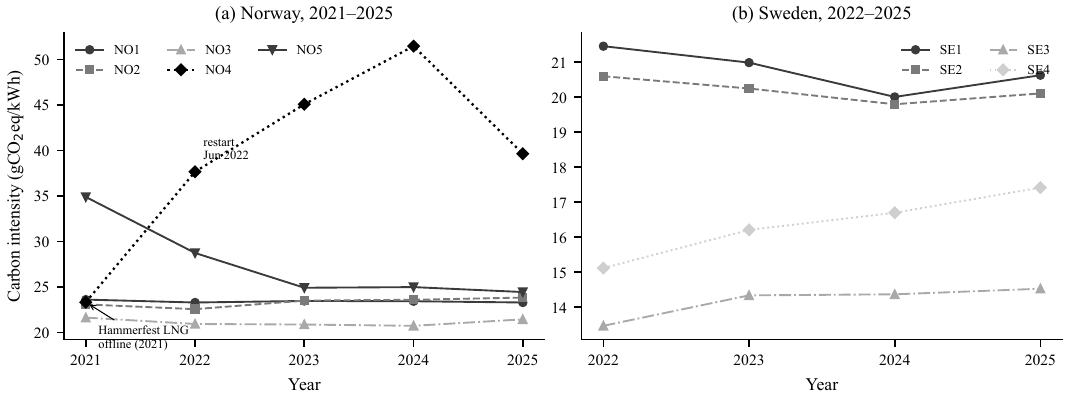}
\caption{Year-over-year carbon-intensity drift, 2021--2025 (Norway) and 2022--2025 (Sweden).}
\label{fig:drift}
\end{figure*}

\begin{figure*}[t]
\centering
\includegraphics[width=0.92\textwidth]{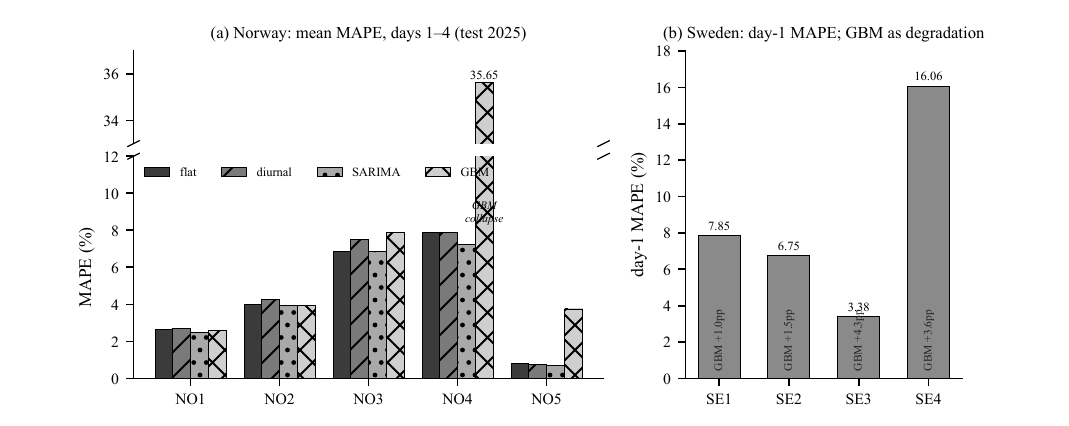}
\caption{Forecast MAPE by zone and model. (a) Norway: mean MAPE, days 1--4, test 2025. (b) Sweden: day-1 MAPE; GBM shown as degradation.}
\label{fig:mape}
\end{figure*}

\section{Discussion}

\subsection{Adoption Pathway}

A dataset that sits in an archive does not close the gap it identifies; the gap closes only when the corrected signal reaches the tools that practitioners already use. We therefore treat adoption as part of the contribution rather than as an afterthought. The per-zone values are contributed upstream into codecarbon, the widely used emissions-accounting library, through two pull requests---\#1260 for the Norwegian zones and \#1262 for the Swedish. The maintainer folded the Swedish values into \#1260 and merged it on 9 September 2026; \#1262 was closed as superseded. The values ship in codecarbon 3.3.1. The framing matters: this is rigor distributed through an existing user base, not a competing platform asking practitioners to switch tools. The contribution replaces codecarbon's production fallback for the Nordic zones---the uniform 18.0 constant---and does not touch its Electricity Maps consumption-based primary path, a scope we state explicitly in the contribution itself. The channel is a realistic one: codecarbon already stores its Nordic factors per zone, reads them dynamically so an update does not break its tests, and has accepted external factor contributions before (the merged precedent PRs \#1039 and \#1224). What rests with the maintainers is the ongoing stewardship of these values---the cadence on which they are refreshed, and the factor table they are kept consistent with---not whether the contribution is accepted. The corrected, source-traceable values are in the tool, through the documented mechanism, with their full provenance attached.

\subsection{From Negative Forecast Result to Update Policy}

The forecast result in Section~III is negative---heavy machine learning does not beat a simple seasonal baseline in either national track---and this is precisely what makes the signal cheap to maintain downstream. Because a SARIMA or persistence baseline suffices, a tool consuming these values does not need a heavy retraining pipeline; a lightweight forecaster, or in the most stable cases no forecaster at all, is enough. The drift characterization turns this into a concrete update policy. The stable zones---NO1, NO2, NO3, and SE2---hold their year-over-year CI within a few percent, so a prior-year figure is a sound proxy and an annual refresh is adequate. The drifting zones require more care, but in different ways that we distinguish: NO5 follows a genuine monotone trend (a real gas phase-out) and so benefits from periodic recomputation, while SE4 crosses the drift threshold on a monotone, solar-driven rise. NO4 is the instructive case---its large movements are an industrial event (the Hammerfest LNG outage and recovery), not a trend, so no update cadence can anticipate them; the honest consequence is that NO4's value must be flagged as period-dependent rather than presented as a stable constant. A downstream integrator can therefore set refresh frequency per zone from the drift evidence, rather than treating all nine zones identically.

\subsection{Limitations}

We state the limitations plainly, because the method's honesty about its own boundary is part of what makes it citable.

The most consequential limitation is that the signal is production-based and not consumption-based. We weight the generation physically produced within a zone; we do not flow-trace imports and exports, and we do not compute marginal emissions. A zone that imports carbon-intensive power therefore has a higher true consumption-based intensity than our production-based figure reflects, and the gap between the two is largest exactly where interconnection is heaviest. Our values should be read as the intensity of in-zone generation, not as the full footprint of consumed electricity. This is a deliberate scope boundary, fixed before computation, not an oversight---but it is a real constraint on how the numbers may be used.

Three narrower limitations follow. First, the forecast tracks the CI level acceptably in all zones but ranks cleaner-versus-dirtier hours poorly in SE1 and SE2, where the concordance index is only 0.581 and 0.596---barely above chance. For scheduling, whose entire purpose is timing, this means the signal supports level-based decisions in those two zones better than it supports fine-grained temporal shifting; we report it as a real limit rather than letting an acceptable MAPE obscure it. Second, the emission factors are IPCC AR5 lifecycle medians taken from a named secondary source, with only the two highest-leverage values (coal and gas) independently confirmed against the IPCC primary document; the remainder rest on the secondary source. Third, Hydro Pumped Storage is assigned the hydro lifecycle factor as a proxy, without tracing the actual charging source, which would in principle shift its footprint. Finally, we restate the cross-track boundary: the Swedish per-zone signal is structurally less distinct than the Norwegian one, with a narrower cross-zone spread; the Swedish contribution is granularity---an archived, recomputable per-zone artefact---and not a stronger or richer signal than Norway's.

\subsection{Future Work}

Several extensions follow naturally, and we describe them as directions rather than promises. The clearest is a consumption-based layer that flow-traces interzonal exchange, which would complement the production-based figures and address the principal limitation above; it is a separate methodological undertaking with its own data requirements. A second is geographic: extending the same method to further Nordic bidding zones, such as the Danish and Finnish zones, conditional on open per-zone generation data of adequate coverage being available for them---a precondition we would verify before claiming a fit, not assume. A third is temporal: a longer drift window as more years accumulate would sharpen the distinction between transient events and durable trends, particularly for the event-driven NO4 zone. None of these is required for the present contribution to stand, and we do not claim capabilities the current method does not have.

\section{Conclusion}

We have presented an open, reproducible, per--bidding-zone carbon-intensity dataset and method for the nine Nordic electricity bidding zones, NO1--NO5 and SE1--SE4---bringing to the bidding-zone level the openness and reproducibility that existing datasets provide only at national resolution. The intensities are derived deterministically from ENTSO-E generation-by-type data and published IPCC AR5 lifecycle factors, and every methodological choice is fixed in a public chain of architecture decision records committed before computation, accompanied by pre-registered drift and forecast analyses.

Three findings carry the work. The uniform placeholder these values replace is wrong in both directions at once---too low where fossil gas is present, too high where nuclear dominates---an error structure that no single scalar can repair and that only per-zone values resolve. The drift analysis separates a genuine generation-mix trend from a one-off industrial event, which matters because the two imply different things about how long a figure stays valid. And the forecast result is honestly negative: heavy machine learning does not beat a simple seasonal baseline, which is not a disappointment but the property that makes the signal cheap to maintain downstream, with a per-zone refresh cadence that follows directly from the drift evidence.

The contribution matters because it is distributed rather than shelved. Instead of standing up a competing platform, we contribute the corrected values upstream into a tool practitioners already use, where they ship with full provenance attached; the artifact is meant to be the citable, source-traceable reference that downstream tools can build on. We are equally clear about the boundary: the signal is production-based and does not compute consumption-based, flow-traced, or marginal emissions, and the Swedish extension is a contribution of granularity---an archived, recomputable per-zone artefact---rather than a signal stronger or more distinct than the Norwegian one.

The durable value, beyond the nine numbers themselves, is the method. Each degree of freedom is locked before the data are seen, and the corrections made along the way remain visible in the project's history rather than being quietly smoothed over. The result is an artifact that can be audited and reproduced, not one that has to be taken on trust---verifiable by construction rather than rationalized after the fact.

\end{document}